# From Checking to Sensemaking: A Caregiver-in-the-Loop Framework for AI-Assisted Task Verification in Dementia Care

Joy Lai[a], Kelly Beaton[b], David Black[b], Bing Ye[a], Alex Mihailidis[a]

[a] University of Toronto, 27 King's College Cir, Toronto, ON M5S 1A1, Canada

[b] Engagement of People with Lived Experience of Dementia Advisory Group and Cross-Cutting Program, Canadian Consortium on Neurodegeneration in Aging, Montreal, Canada.

## Abstract

Informal caregivers play a central role in enabling people living with dementia (PLwD) to remain at home, yet they face persistent challenges verifying whether daily tasks have been completed. Existing digital reminder systems prompt actions but rarely confirm outcomes, leaving caregivers to double-check tasks manually. This study explores how generative artificial intelligence (AI) might support caregiver-led task verification without displacing human judgment. We combined qualitative interviews with ten caregivers and one PLwD with a speculative simulation probe using a generative large language model to generate follow-up questions and flag responses for verification. Using template analysis, we identified three interrelated patterns of reasoning: detecting anomalies, constructing trustworthy evidence, and calibrating trust and control. These insights informed the Caregiver-in-the-Loop Task Verification (CLTV) framework, which models verification as a collaborative cycle of anomaly detection, evidence triangulation, AI-assisted summarization, and accountability circulation centered on caregiver oversight. CLTV advances human-AI collaboration theory by situating interpretability, trust, and control within the relational and emotional realities of dementia care and by offering design principles for transparent, adjustable, and context-aware AI support. We contribute a care-centered extension of human-AI collaboration theory, demonstrating how interpretability and trust can be operationalized through caregiver oversight.

## 1 Introduction

Home-based care enables people living with dementia (PLwD) to remain in familiar environments and preserve autonomy for as long as possible. Yet this model depends heavily on informal caregivers, often family members, who must organize routines, monitor safety, and confirm that essential tasks are completed (Armstrong, 2022; Curnow et al., 2021; Frias et al., 2020; Zwerling et al., 2016). Despite many digital reminder tools designed to support these routines (Jönsson et al., 2019; McGoldrick et al., 2021; Mettouris et al., 2023; Peres and Campos, 2023), caregivers report a persistent "verification gap": reminders prompt actions but rarely confirm outcomes. As a result, caregivers often have to double-check manually by calling, visiting, or visually confirming completion, which undermines the promise of automation. Addressing this gap requires systems that not only remind but also help establish trustworthy evidence of task adherence.

HCI research has long examined how technologies are adopted, resisted, and emotionally interpreted in dementia care and aging-in-place contexts (Lazar et al., 2017; Lindsay et al., 2012). Studies highlight tensions between independence, safety, and feelings of surveillance; the need for technologies that support dignity and autonomy; and the relational nature of caregiving within the home (Berridge et al., 2021; Brookman et al., 2023; Peres and Campos, 2023). This body of work positions task verification not merely as a technical challenge but as a sociotechnical one involving routines, relationships, and moral responsibility (Suchman, 2006).

Recent advances in generative artificial intelligence (AI), particularly large language models such as GPT-4, introduce new possibilities for addressing this problem. Rather than relying on sensors or manual reporting, language models could generate *follow-up questions* after a reminder ("Is your toothbrush wet?" rather than "Did you brush your teeth?"), interpret responses, and flag potential concerns for caregiver review (Chen et al., 2024; Hossain et al., 2024; Zaman et al., 2023). Such an approach may offer a lightweight, privacy-preserving alternative to sensor-based activity recognition (Lai et al., 2025). However, the introduction of AI into caregiving decision-making raises significant questions about trust, interpretability, and the caregiver's role in verification (Andargoli et al., 2024; Berridge et al., 2021). If AI systems take on aspects of observation or judgment, how can caregivers remain "in the loop" and maintain oversight of what the system infers or decides?

Within HCI and human-computer studies, human-AI collaboration research has shown that trust, interpretability, and control must be negotiated dynamically as users make sense of AI suggestions (Ehsan et al., 2021; Liao et al., 2020). Work on mixed-initiative systems, conditional delegation, and trust calibration emphasizes that effective collaboration requires transparency about how inferences are generated and mechanisms for users to adjust system behaviour (Amershi et al., 2019; Lai et al., 2022). Extending these insights to informal care raises new questions about how non-expert caregivers understand, contest, or rely on AI-generated cues in emotionally charged domestic environments (Eiband et al., 2018).

However, while human-AI collaboration frameworks such as human-in-the-loop or mixed-initiative systems have been widely applied in professional domains like healthcare diagnostics or content moderation, they remain under-theorized for informal care (Amershi et al., 2019; Lai et al., 2022; Shneiderman, 2020). Existing frameworks typically assume expert users who operate within formal decision boundaries, whereas informal caregivers engage in emotionally charged, context-rich reasoning that blends interpretation, empathy, and risk management (Berridge et al., 2021; Brookman et al., 2023; Lazar et al., 2017). This theoretical gap limits our understanding of how interpretability, trust, and control should be designed when AI participates in domestic and relational forms of care.

To explore this question, we conducted a qualitative design study that combines caregiver interviews with a *speculative simulation probe*. The study examined how caregivers conceptualize task verification, how they evaluate the usefulness and risks of AI-generated follow-up prompts, and what mechanisms they envision for maintaining control over such systems. The simulation used authentic reminder data to illustrate a prototype AI pipeline that generates, analyzes, and flags responses to follow-up questions. These examples served as conversational prompts during interviews, helping participants articulate design values such as transparency, editability, and emotional appropriateness. This dual method allowed us to move beyond technical feasibility, focusing instead on how caregivers conceptualize verification as an interpretive and moral process rather than a computational one.

Through **template analysis** of the interview data and reflection on the simulation outputs, we identified three interrelated patterns of caregiver reasoning: (1) *detecting and interpreting anomalies*, (2) *constructing trustworthy evidence*, and (3) *calibrating trust and control*. These patterns form the foundation of the **Caregiver-in-the-Loop Task Verification (CLTV)** framework, which conceptualizes task verification as a collaborative process between humans and AI.

The study makes three primary contributions:

1. **Empirical insights** into caregivers' verification practices and their expectations of AI involvement in daily routines.
2. **A formative simulation probe**, using real reminder data, that illustrates, rather than evaluates, how context and caregiver feedback affect AI-generated verification prompts.
3. **The CLTV framework**, which translates caregiver reasoning into a design model for transparent and adjustable AI-supported task verification.

By integrating qualitative inquiry with speculative prototyping, this study advances understanding of how human-AI collaboration unfolds in domestic care. We contribute a care-centered extension of human-AI collaboration theory, demonstrating how interpretability and trust can be operationalized through caregiver oversight. The proposed Caregiver-in-the-Loop Task Verification (CLTV) framework reframes verification as a process of relational sensemaking between caregivers and AI systems, offering practical design principles for transparent, adjustable, and context-aware automation in dementia care.

The primary research question guiding this work is: How can AI participate in task verification for dementia care while preserving caregiver interpretive control and accountability?

# 2 Related Work

This study builds on research across three intersecting areas: (1) task adherence and verification in dementia care, (2) human-AI collaboration and explainability in assistive technologies, and (3) the emerging use of generative AI in cognitive and caregiving contexts. Together, these literatures reveal the persistent challenges of confirming task completion in home-based dementia care and the need for AI systems that are both interpretable and adjustable by caregivers.

## 2.1 Task Adherence and Verification in Dementia Care

Home-based care for PLwD depends on consistent support for activities of daily living, such as hygiene, meals, medication, and appointments. These tasks often become fragmented as memory and executive function decline. Digital reminder systems, such as *eSticky* and *MindMate*, have been developed to cue PLwD through visual or auditory prompts that reduce the need for caregiver intervention (Jönsson et al., 2019; McGoldrick et al., 2021; Mettouris et al., 2023). While these systems can improve task initiation, their effectiveness is constrained by what caregivers call *the verification problem*: even if a PLwD acknowledges a prompt, there is rarely any confirmation that the task was completed (McGoldrick et al., 2021; Sanchez et al., 2024b).

HCI scholarship on dementia technologies similarly highlights the challenges of balancing independence, autonomy, and oversight in the home (Lazar et al., 2017; Lindsay et al., 2012). Prior work has shown that technologies intended to support safety or adherence often introduce concerns around surveillance, intrusion, and loss of dignity, requiring careful design to negotiate visibility and control (Berridge et al., 2021; Brookman et al., 2023). These studies position reminder and monitoring systems as deeply relational artifacts that shape how caregiving responsibilities are coordinated and understood (Mort et al., 2013; Procter et al., 2016).

Current verification strategies often rely on direct observation, follow-up calls, or sensor-based monitoring (Civitarese, 2019; Das et al., 2023; Hao et al., 2016; Lai et al., 2025; Lobanova et al., 2023; Na et al., 2012; Shahid et al., 2023; Sim et al., 2011; Simonsson et al., 2023). While these technologies can infer likely task completion (for example, detecting cooking activity through appliance use), they introduce issues of cost, intrusiveness, and privacy, and still require caregivers to interpret what actually happened. As a result, many revert to manual checking, which reintroduces stress and undermines automation's benefits.

Several studies have emphasized that effective verification relies on contextual reasoning, which involves understanding not only whether an action occurred but also whether it took place in a meaningful or safe way (Michalakis et al., 2021; Schmidtke Hedda R., 2025). For instance, taking medication is different from taking the correct medication at the correct time; eating lunch differs from eating enough food. These nuances underscore that verification is not simply an *act of detection* but a *process of interpretation*. This distinction motivates approaches that support caregiver sensemaking rather than replace it.

### 2.2 Human-AI Collaboration and Explainability in Assistive Technologies

Human-AI collaboration frameworks within assistive and healthcare technologies highlight the need for systems that distribute cognitive work between humans and machines, maintaining caregiver oversight while reducing workload (Amershi et al., 2019; Shneiderman, 2020). *Human-in-the-loop* paradigms conceptualize AI as a supportive agent that augments decision-making rather than fully automates it (Tsiakas and Murray-Rust, 2022). In dementia care, this balance is particularly delicate: caregivers must trust the system's suggestions without ceding control over sensitive care decisions. In our study, these principles manifest concretely as editable phrasing for follow-up questions and adjustable alert thresholds, giving caregivers fine-grained agency over how the system reasons and when it escalates.

In HCI and human-computer studies, research on explainable AI and trust calibration demonstrates that users' willingness to rely on AI systems hinges on their ability to understand, interrogate, and adjust system outputs (Liao et al., 2020). Explanations function not only as cognitive aids but as interactional resources through which people assess the appropriateness of algorithmic suggestions. Studies show that when users understand how an AI arrives at its suggestions, whether through transparent reasoning or adjustable parameters, they are more likely to adopt its recommendations responsibly (Eiband et al., 2018; Miller, 2019). Conversely, opaque automation can erode confidence, leading users either to disregard alerts or to over-rely on them without critical assessment. In the caregiving context, this translates to a need for interpretable verification cues, where AI recommendations can be inspected, edited, or overridden by caregivers.

Mixed-initiative HCI models further suggest that users benefit from systems that support shared control, allowing humans to steer, edit, or override system behaviours as circumstances evolve (Amershi et al., 2019; Eric Horvitz, 1999). A caregiver-in-the-loop paradigm naturally extends these ideas to domestic care, where caregivers must constantly calibrate trust and responsibility across people, routines, and technologies (Lai et al., 2022; Procter et al., 2016).

HCI research also underscores the *emotional dimension* of assistive interaction (Lazar et al., 2017). Caregiving technologies must manage not only information flow but also affective tone, avoiding prompts that cause frustration, embarrassment, or perceived surveillance (Brookman et al., 2023; Cannings et al., 2024). Caregivers value systems that "speak in their voice," meaning those that reflect their caregiving style, priorities, and sense of empathy (Brookman et al., 2023; Sanchez et al., 2024b). These findings suggest that AI systems for dementia care should be designed for co-adaptation: they learn from caregiver input, but also provide outputs that caregivers can meaningfully adjust (Tsiakas and Murray-Rust, 2022).

This literature motivates the concept of caregiver-in-the-loop design, where caregivers retain authority over what the system infers, how prompts are phrased, and when alerts are triggered. Our study extends this line of work by examining how caregivers themselves conceptualize verification, what they consider trustworthy evidence, and how they want control distributed between themselves and an AI system.

### 2.3 Generative AI for Cognitive Support and Caregiving

Generative AI models, such as GPT-4, have recently been explored as conversational companions or assistants for individuals with cognitive impairment. Early work suggests their potential to provide reminders, suggest activities, and simulate empathetic dialogue (Chen et al., 2024; Gilman et al., 2024; Hossain et al., 2024). For example, Chen et al. (Chen et al., 2024) demonstrated a ChatGPT-enabled assistive platform that supports activities of daily living for people with mild dementia, while Zaman et al. (Zaman et al., 2023) explored the use of GPT-based assistants for caregiver education and emotional support. These studies emphasize the potential of large language models (LLMs) to personalize assistance through natural language.

However, the use of generative AI in caregiving remains largely *speculative*. Key concerns include bias in generated content, difficulty maintaining consistent tone, and limited capacity for contextual grounding (Afreen et al., 2025). LLMs often struggle to infer whether an event actually occurred, which is an issue central to task verification. Furthermore, few systems integrate caregiver feedback into the AI's generative process. Most current prototypes focus on direct interaction with PLwD rather than collaborative caregiver-AI workflows (Chen et al., 2024; Gilman et al., 2024; Hossain et al., 2024).

This emerging line of work builds on longstanding HCI interest in conversational interfaces and naturalistic interactions for supporting older adults (Koo and Vizer, 2019; Pradhan et al., 2018). Prior HCI research has examined how conversational agents can scaffold memory, routine, and emotional wellbeing (Ammari et al., 2019), while also highlighting the risks of ambiguity, unpredictability, and inconsistent tone (Luger and Sellen, 2016; Porcheron et al., 2018). Situating generative AI within this tradition emphasizes the need for transparency, editability, and contextual sensitivity when deploying open-ended language models in caregiving contexts.

At the same time, emerging research argues that LLMs should be used not as autonomous decision-makers but as interactive design materials that help surface human reasoning and support co-design (Mao et al., 2023). In this spirit, our study treats GPT-4 not as a deployable solution but as a *speculative probe* to explore how AI could participate in verification without displacing caregiver judgment (Zhou and Chen, 2025). By generating examples of potential AI follow-up questions and response categorizations, the simulation provided a way to provoke reflection and reveal caregivers' design expectations.

### 2.4 Frameworks for Human-AI Collaboration in Care Contexts

Several conceptual frameworks describe how humans and AI systems interact in complex domains. Classic models such as Shneiderman's Human-Centered AI framework emphasize agency, transparency, and user control (Shneiderman, 2020), while newer approaches like Conditional Delegation (Lai et al., 2022) define boundaries where AI can act autonomously or defer to humans. More broadly, reviews of human-AI collaboration show that joint systems should enable mutual adaptation, feedback loops, and dynamic task allocation (Amori Molina and Kharlov, 2024; Puerta-Beldarrain et al., 2025).

However, many of these frameworks are grounded in interactions between AI and expert users (e.g. clinicians, analysts) rather than informal or family caregivers. Caregiving in home settings involves emotional, relational, and contextual reasoning that goes beyond procedural decision support. In dementia care specifically, few models address how caregivers interpret anomalies, integrate subjective knowledge, or calibrate trust over time.

Recent HCI research offers a complementary lens by emphasizing that collaboration with AI systems is not merely a matter of task allocation but an ongoing process of negotiation in which users interpret, challenge, and reinterpret system outputs (Bansal et al., 2019; Lindgren, 2025). This has led to formulations of co-adaptive systems in which both users and technologies adjust to one another over time (Zhao et al., 2025). Applying this perspective to caregiving foregrounds how caregivers develop tacit routines, emotional attunement, and interpretive heuristics that may not align with traditional expert-centric notions of decision support.

The **Caregiver-in-the-Loop Task Verification (CLTV)** framework we propose responds to this limitation. It adapts human-AI collaboration principles to the domestic caregiving environment by embedding mechanisms such as editable prompts, adjustable alert thresholds, and transparent flagging, all maintained under caregiver oversight. In contrast with many generic models, CLTV centers the interpretive and affective judgments caregivers make, treating AI as a supporting partner rather than authority.

### 2.5 Summary and Research Gap

Across these bodies of work, four interrelated gaps persist.

1. **Verification remains technologically fragmented.** Existing reminder and monitoring systems address task initiation but rarely integrate mechanisms for confirming completion, leaving caregivers to rely on manual checking.
2. **Automation raises challenges of explainability and control.** Although AI and sensor-based systems can assist with detection, caregivers often lack insight into how the system reaches its conclusions or the ability to adjust its reasoning.
3. **Generative AI for caregiving is largely unexplored.** Early prototypes demonstrate feasibility but provide little evidence of how such systems might integrate into everyday caregiving routines or accommodate caregiver oversight.
4. **Human-AI collaboration frameworks overlook relational caregiving.** Existing models emphasize professional decision-making and technical accountability, but few capture the emotional, contextual, and moral reasoning that shape informal caregiving in the home.

These gaps highlight a need for design frameworks that articulate *how caregivers reason about verification* and *how AI systems can participate in that reasoning without displacing it*. This study responds to that need by combining qualitative interviews with caregivers and a speculative simulation probe to examine the interpretive, emotional, and procedural aspects of task verification. From this inquiry, we develop the CLTV framework, a conceptual model that translates caregivers' verification practices into design principles for transparent, adjustable, and context-aware AI support in dementia care.

In contrast with existing human-AI frameworks, the CLTV framework extends collaboration theory into the domestic and relational domain of dementia care. Whereas these established approaches emphasize reliability, bounded autonomy, and procedural control for expert users, CLTV foregrounds interpretive reasoning, emotional labor, and shared accountability among non-expert caregivers. By situating interpretability and trust calibration within everyday care routines, CLTV reframes transparency not merely as model explanation but as an ongoing dialogue between human judgment and AI-generated cues. This care-centered orientation fills a critical theoretical gap in current human-AI collaboration research and provides design heuristics for adaptive, emotionally attuned systems.

# 3 Methods

## 3.1 Study Overview

This study combined qualitative interviews with a **simulation-based speculative probe** to explore how caregivers conceptualize task verification and how generative AI might support this process. The study was grounded in a **Caregiver-in-the-Loop** approach, positioning caregivers not as passive recipients of technology but as co-interpreters and co-designers of verification logic. The investigation unfolded in two complementary phases:

1. **Simulation Probe Development and Mini Test** - conducted collaboratively with lived-experience advisors to examine the feasibility of using generative AI (GPT-4) to create follow-up questions and flag responses related to digital reminders.
2. **Qualitative Interviews** - with ten informal caregivers and one PLwD to explore expectations, design preferences, and perceptions of AI-mediated verification.

The interviews reported in this paper formed part of a broader exploratory study on designing enhanced digital reminder systems for dementia care. Participants discussed a range of potential features including anomaly detection, task verification, event logging, communication within care networks, and alert customization. The present analysis focuses specifically on the dimension of **task verification**, identified early in the analysis as a critical yet underexplored challenge in caregivers' management of daily routines. This thematic subset was chosen because it most directly informed the design of the Caregiver-in-the-Loop Task Verification (CLTV) framework described in later sections.

## 3.2 Participants

### 3.2.1 Caregivers and PLwD

Ten informal caregivers of PLwD (4 female, 6 male) participated in semi-structured interviews. Participants were recruited through caregiver networks and dementia research registries in Canada. All had direct experience supporting a family member with early to moderate dementia in a home-based setting. One individual living with early-stage dementia also participated to provide complementary perspective on reminder adherence and autonomy. Caregiver demographics are shown in Table 1.

Participants were asked to reflect on potential features that could enhance the reminder system, including verification of task completion, coordination among caregivers, and options for customizing alerts. Discussion of task verification provided particularly rich insight into caregivers' interpretive reasoning and forms the analytic focus of this paper. The intended scope of the design exploration is early to moderate dementia, where PLwD can still engage with prompts and provide brief responses.

*Table 1 Participant demographics and caregiving situation*

| Relationship | Gender | Age range | Living arrangement | Care duration | Care intensity | Caregiver Support | Tech familiarity |
|---|---|---|---|---|---|---|---|
| **1** | Adult grandchild | M | 18-29 | Co-residing | 1-3 yrs | Sole caregiver | Moderate |
| **2** | Spouse | F | 60+ | Co-residing | 4-6 yrs | Sole caregiver | Low |
| **3** | Adult child | F | 40-50 | Separate | 6+ yrs | Shared (Personal Support Worker) | Moderate |
| **4** | Adult grandchild | M | 30-39 | Co-residing | 1-3 yrs | Shared (family) | Moderate |
| **5** | Spouse | M | 60+ | Co-residing | 6+ yrs | Sole caregiver | Low |
| **6** | Adult child | F | 30-39 | Co-residing | 6+ yrs | Shared (family) | High |
| **7** | Adult child | M | 40-50 | Separate | 4-6 yrs | Shared (sibling) | High |
| **8** | Adult grandchild | F | 18-29 | Co-residing | 1-3 yrs | Shared (family) | Moderate |
| **9** | Adult grandchild | M | 18-29 | Co-residing | 1-3 yrs | Shared (family) | High |
| **10** | Spouse | M | 60+ | Co-residing | 6+ yrs | Sole caregiver | Moderate |

#### 3.2.2 Recruitment and Eligibility

Participants were recruited via caregiver networks and dementia organizations in Canada, including the Alzheimer Society of Canada, through postings and newsletters. Eligibility required: (a) age ≥18; (b) current or recent informal caregiving for a family member or friend with dementia living at home; (c) ability to participate in English via videoconference; and (d) willingness to discuss reminder/verification practices. Exclusion criteria included exclusively institutional care contexts and primary clinical/paid caregiving roles. Participants received $30 CAD honoraria.

#### 3.2.3 Lived-Experience Advisors

In addition to interview participants, two long-term Engagement of People with Lived Experience of Dementia (EPLED) advisors served as co-design partners throughout the study (Snowball et al., 2024). Both were experienced caregivers who regularly collaborate on dementia research initiatives. Their contributions shaped the simulation workflow, prompting design, and evaluation logic. EPLED advisors reviewed coding templates for resonance and co-designed prompts but did not code transcripts or participate as interviewees.

### 3.3 Simulation Probe Design

The simulation probe was included to explore how generative AI could serve as a reflective design tool rather than a performance benchmark. Its purpose was to prompt discussion and reveal caregivers' values around verification, not to evaluate technical accuracy. The probe was developed to explore the feasibility and design implications of integrating generative AI into a digital reminder system for PLwD. It was not intended as a fully deployable prototype, but rather as a formative design artifact to provoke reflection among caregivers.

#### 3.3.1 Dataset

The simulation used an anonymized historical dataset of 64 digital reminders drawn from two previous home-based usability studies with caregiver-PLwD dyads (Sanchez et al., 2024a). The reminders reflected typical caregiving scenarios, including appointments, meals, hygiene tasks, social activities, and safety checks. Medication-related reminders were excluded to avoid ethical concerns associated with health-related decision automation. All identifiers were removed, and contextual details (e.g., household routines, caregiver relationship, task priority) were retained only in abstracted form.

#### 3.3.2 Simulation Loop

The simulation represented an idealized **AI-assisted task verification loop** (Figure 1). For each reminder, the system generated:

- One **AI-created follow-up question**,
- Three **simulated responses** representing *low*, *medium*, and *high* concern levels.

The generative model (GPT-4) was prompted using few-shot examples co-designed with EPLED advisors. The prompting logic incorporated task metadata such as type ("meal", "appointment"), priority level, and contextual notes (e.g., "Sam is the household dog"). These inputs guided GPT-4 to generate concise, observation-based follow-up questions suitable for early-stage dementia communication (e.g., "Is your breakfast plate empty?" instead of "Did you eat breakfast?").

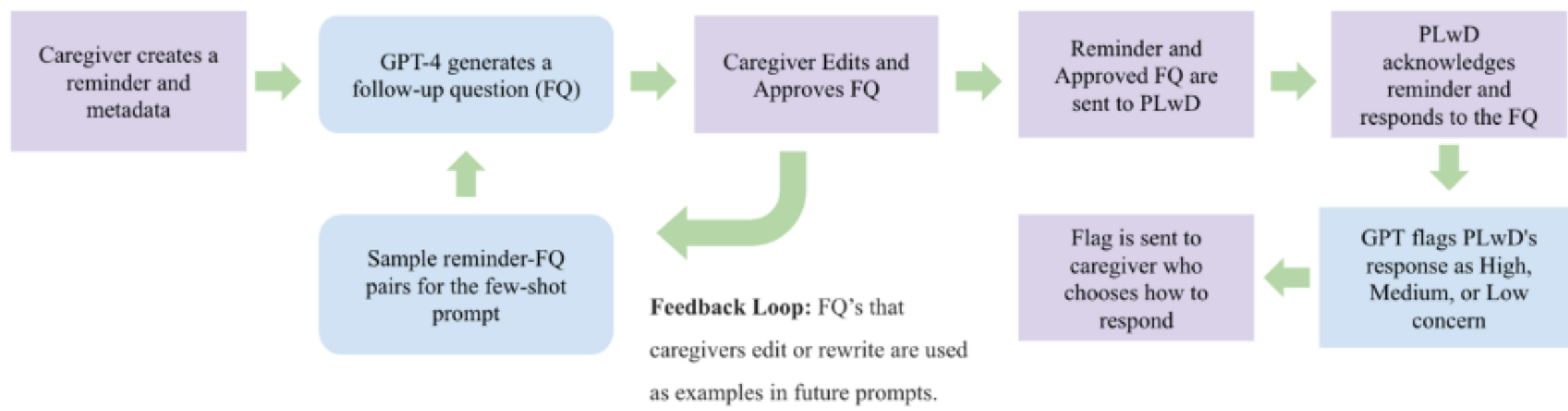

*Figure 1 Theoretical pipeline for the idealized AI-assisted task verification loop*

#### 3.3.3 Caregiver Feedback Integration

The advisors helped design a **caregiver feedback loop** within the simulation: caregivers could review each generated question, edit or rewrite it, and the modified examples were reintroduced into the few-shot prompt set. This iterative refinement tested whether incorporating caregiver-authored examples improved subsequent question quality and contextual appropriateness.

#### 3.3.4 Mini-Test Evaluation

To assess feasibility, a small-scale evaluation compared GPT-4 outputs generated with versus without contextual caregiver input. Each of the 64 reminders produced one follow-up question under both conditions. The resulting 128 questions were evaluated using a structured 12-item Follow-up Question Quality Checklist co-created with EPLED advisors, emphasizing clarity, contextual relevance, memory independence, and supportive tone.

For each follow-up question, GPT-4 also produced three simulated responses corresponding to *low*, *medium*, and *high* concern. Responses were categorized according to a decision tree based on task criticality and response clarity. This allowed examination of whether context improved GPT-4's accuracy in flagging responses appropriately. The evaluation results (Section 4) informed later interview discussions about how caregivers interpret "good" follow-up questions and desired levels of automation. The simulation thus functioned as a *design probe* (Boehner et al., 2007), enabling participants to articulate values, boundaries, and expectations around AI-assisted verification through concrete examples. Results are descriptive and illustrative rather than inferential; the probe functions as a design artifact to elicit values and feasibility considerations, not as a statistical evaluation.

The 12-item Follow-up Question Quality Checklist was applied by the first author. EPLED advisors reviewed the instrument and spot-checked a purposive subset (≈10%) of examples for face validity; disagreements about item interpretation were resolved through discussion. Given the probe's formative intent, we report descriptive counts/percentages rather than inferential statistics. The checklist and decision tree are provided in Appendix A and Appendix B.

A concise summary table contrasting outputs with vs. without caregiver-provided context (clarity, contextuality, memory-light phrasing, tone) is presented in Table 2.

*Table 2 Examples contrasting GPT-4 follow-up questions generated with vs. without caregiver-provided contextual information.*

| Reminder | Original AI Follow-up (No Context) | Revised AI Follow-up (With Caregiver Context) |
|---|---|---|
| **[Redacted] coming tomorrow** | Will you set out fresh towels for [Redacted]'s visit? | Are you excited for [Redacted]'s visit tomorrow |
| **Home and Garden show** | Did you check the schedule for the Home and Garden show? | Are you watching the Home and Garden Show on TV? |
| **Leave Cambridge** | Are you ready to head out for your activity? | Have you packed your bag to leave Cambridge? |
| **Brunch** | Did you set the table for brunch? | Did you enjoy your brunch? |

## 3.4 Interview Design and Data Collection

### 3.4.1 Purpose and Scope

Interviews were conducted to understand caregivers' interpretations of task verification, their comfort with AI-generated follow-up questions, and their expectations for oversight and control. The simulation outputs served as *conversation prompts*, allowing participants to react to concrete examples of how an AI system might verify reminders.

### 3.4.2 Procedure

Each participant took part in a one-hour semi-structured interview conducted via Zoom between June-October of 2025. Interviews were audio-recorded with consent and auto-transcribed using Zoom's built-in transcription feature, with manual correction for accuracy.

Participants were first asked about their current reminder and monitoring practices (e.g., "How do you usually check that tasks have been completed?"). They were then shown examples of AI-generated follow-up questions and flagging outputs from the simulation and asked to reflect on their perceived usefulness, tone, and trustworthiness. Questions probed themes such as:

- What kinds of follow-up questions feel supportive or intrusive?
- How much control should caregivers retain over question phrasing or thresholds?
- In what situations would they trust an AI system to flag missed tasks?

Interviews concluded with open discussion of potential benefits, risks, and ethical considerations of integrating AI verification into daily caregiving. We did not collect quantitative ratings; the probe outputs served as discussion prompts to surface design expectations and concerns.

### 3.5 Data Analysis

Interview transcripts were analyzed using Template Analysis (King, 2012), a structured yet flexible approach suited to applied HCI and health-technology research. This method involves developing a hierarchical "template" of codes and themes that can be iteratively refined.

#### 3.5.1 Coding Procedure

The first author conducted all coding using Taguette, an open-source tool for qualitative data analysis.

- **Step 1:** Initial inductive codes were generated from the transcripts, capturing caregivers' actions, interpretations, and reactions to AI-mediated verification.
- **Step 2:** Codes were clustered into higher-order categories representing different aspects of verification reasoning (e.g., "noticing deviations," "cross-checking," "judging reliability").
- **Step 3:** These clusters were organized into an evolving template, refined through iterative review with the two EPLED advisors, who provided interpretive feedback based on their lived experience.
- **Step 4:** The final template comprised three overarching themes: *Detecting and Interpreting Anomalies*, *Constructing Trustworthy Evidence*, and *Calibrating Trust and Control*. Each contains multiple subthemes that mapped the practical and emotional dimensions of caregiver decision-making.

The analysis began with 35 initial inductive codes, which were iteratively consolidated into 8 second-level categories and three top-level themes. We monitored meaning saturation around verification; by interview 8, new data elaborated existing codes without introducing substantively new categories, supporting adequacy for the present analysis. An audit trail documented code merges/splits and template revisions.

Although the overall template included themes related to other potential system features (e.g., event tracking, coordination tools, and alert personalization), the analysis presented here focuses on the subset of themes concerned with task verification and oversight. These themes represented a coherent conceptual strand within the broader dataset and were most relevant to the design problem addressed in this paper.

#### 3.5.2 Integration with Simulation Insights

Findings from the simulation probe were not treated as qualitative data but used analytically to contextualize interview responses. For example, when caregivers discussed question phrasing, comparisons were drawn to GPT-4-generated examples to explore where automation aligned or diverged from caregiver reasoning. This integrated analysis linked empirical caregiver perspectives with the speculative design implications of the AI prototype.

#### 3.5.3 Trustworthiness and Quality Criteria

We enhanced credibility through reflexive memoing, periodic peer debriefs with EPLED advisors, and member-style checking of emerging interpretations during advisor sessions. Dependability was supported by an audit trail of codebook and template changes (dates, rationales). Transferability was addressed via thick description of caregiving contexts and tasks. Given the single-coder approach, we did not compute inter-coder reliability; instead, we used advisor reviews of the coding template to assess resonance and clarity.

### 3.6 Researcher Positionality and Reflexivity

The first author conducted all interviews, coded the data, and developed the simulation prototype. Their background in biomedical engineering and human-computer interaction informed a technical understanding of AI systems but not direct caregiving experience. Reflexivity was maintained through analytic memoing and periodic debriefs with EPLED advisors to check interpretive resonance and to ensure that emergent themes reflected caregiver realities rather than researcher assumptions. The author's dual role as developer and analyst was explicitly documented to promote transparency about potential bias and interpretation boundaries. The author engaged in bracketing memos prior to interviews and after each coding session to surface assumptions related to technical feasibility and caregiving norms.

### 3.7 Ethical Considerations

The study received ethics approval from the University of Toronto Research Ethics Board. All participants provided informed consent prior to participation and were assured of confidentiality and the right to withdraw at any point. The historical reminder dataset used for the simulation was fully anonymized, with all identifying information removed. No medical advice or diagnostic functions were included in the AI simulation. Medication-related reminders were excluded to prevent unintended inference about clinical behavior. Data were stored securely on encrypted institutional servers, and all quotes presented in results are anonymized to protect participant identity.

### 3.8 Summary

By combining a speculative simulation probe with qualitative interviews analyzed through Template Analysis, this study bridges conceptual and technical inquiry. The simulation provided a structured artifact to elicit caregivers' design values, while the interviews revealed the interpretive and emotional dimensions of verification. Together, they informed the development of the **CLTV framework**, which operationalizes how caregivers wish to remain "in the loop" within AI-supported dementia care systems.

## 4 Results

### 4.1 Overview

Caregivers conceptualized task verification not as a binary "done/not done" check, but as a dynamic, context-dependent process involving interpretation, evidence, and emotional judgment. Across both the interviews and the simulation probe, verification emerged as an iterative cycle of noticing, interpreting, and confirming. The simulation results provided early indications of how generative AI could support this process by producing contextually sensitive follow-up questions and prioritizing anomalies, while the interviews revealed how caregivers made sense of these verification cues in practice. Three interrelated themes capture how caregivers envisioned collaboration with AI systems in dementia care:

1. **Anomalies as verification signals**,

2. **Constructing trustworthy evidence**, and
3. **Conditional trust in AI and automation**.

Together, these themes describe a *care-centered ecology of verification*, in which AI may scaffold human reasoning but caregivers remain the arbiters of what constitutes reliable proof. These findings address our central research question of how AI can participate in task verification while preserving caregiver interpretive control and accountability, showing that verification in practice is negotiated through anomaly detection, evidence construction, and calibrated trust.

### 4.2 Theme 1: Anomalies as Verification Signals

The simulation probe demonstrated that contextual information improved the AI's ability to generate relevant follow-up questions and to classify responses by urgency (Section 3.3.4). When shown examples of these outputs, caregivers often described them as helpful "starting points" for detecting anomalies, or deviations from expected routines that might indicate risk. However, they consistently reframed verification not as anomaly *detection* but as *interpretation*: noticing patterns, judging their meaning, and deciding whether they merited intervention.

*"I don't care if she waters the plants, but I do care if she takes her medication, I do care if she is leaving the house, I do care if she turned off the stove."*

Caregivers distinguished between signal and noise. Some anomalies—like skipped meals or an unlatched door—triggered concern; others, such as inconsistent minor routines, were treated as normal variation. The need to prioritize anomalies was a recurrent theme: systems that surfaced every deviation risked overwhelming users and amplifying stress. As one participant noted, caregivers wanted "just the things that matter…meals, meds, anything unusual." Caregivers repeatedly prioritized anomalies around medications and meals as verification triggers: "I said the top priority is medications." Another added, "Mealtimes… that's still at the first after medications." Some valued actionable summaries that feed directly into clinical encounters: "Doctors are asking, how often is she taking her medications? … a summary report would be helpful."

In this sense, anomalies were *entry points for inquiry*, not evidence in themselves. Caregivers emphasized that missing data or irregular behavior did not automatically imply failure. Instead, anomalies prompted them to gather more information from other cues, such as objects, routines, or interpersonal check-ins. This pattern mirrored the simulation's results: AI handled high-stakes tasks more clearly than routine ones, reinforcing caregivers' view that anomalies require human interpretation. They also monitored subtle signs of disengagement as early indicators, such as shorter replies or reduced interaction that might mask decline. One caregiver explained, "If there is a shift and a change… sometimes they're very good at masking."

**Design implication:** AI systems should surface deviations as *contextualized stories* rather than raw alerts, helping caregivers distinguish between routine variability and genuine concern. In the CLTV framework, this insight underpins the principle of context-sensitive anomaly surfacing, emphasizing interpretability over frequency.

These interpretive moves naturally lead into how caregivers construct and cross-check evidence once an anomaly draws their attention, which is explored in the next theme.

## 4.3 Theme 2: Constructing Trustworthy Evidence

For caregivers, verification was a process of assembling **trustworthy evidence** across incomplete, inconsistent, or ambiguous sources. The simulation probe suggested that follow-up questions grounded in concrete, perceptual cues ("Is your toothbrush wet?") were rated higher than memory-based ones ("Did you brush your teeth?"). When caregivers discussed these examples, they often described using similar phrasing strategies based on their experience, relying on observation, simplicity, and sensory checks.

### 4.3.1 Simplified, Memory-Light Questioning

Caregivers highlighted that PLwD often respond reflexively to memory-based questions, producing false positives ("They'll always say yes"). Caregivers favored observable, concrete phrasing over memory checks: "Did you use hot or cold water to wet your toothbrush?" and "If I ask her to show me something rather than tell me, there's less pressure to remember."

This aligns with the simulation: GPT-4 generated better follow-up questions when prompted with caregiver-style examples, showing how caregiver phrasing can iteratively train the AI to communicate appropriately.

### 4.3.2 Material and Environmental Traces

When verbal confirmation failed, caregivers sought physical or environmental indicators of task completion.

*"I could tell by a wet towel… later in the stages, I'd have to go look for evidence. Did she eat? Is the microwave still warm?"*

*"She said she bathed her first day, bathed her second day… I come home… there's no towels to be done… this doesn't match up."*

These tangible "evidentiary traces" served as embodied verification, representing an interpretive reconstruction of what happened rather than a direct observation. Caregivers cross-checked the environment to infer completion, blending sensory data with contextual knowledge ("She said she bathed, but the towels are dry").

AI systems that aim to support verification should mirror this evidential reasoning. Rather than asserting completion, they should show *why* a task appears complete, making their logic inspectable. In CLTV, this insight translates into the design principle of evidence triangulation, which integrates conversational, contextual, and material cues into interpretable summaries.

### 4.3.3 Shared Records and Collective Accountability

Verification also operated socially. In multi-caregiver contexts, participants used whiteboards, logbooks, or group texts to coordinate daily routines. Yet these records were often inconsistent or incomplete, generating skepticism toward "official" logs that lacked material corroboration.

*"[Personal Support Worker] was literally just almost copying and pasting the same notes to save her the time… the only one I can trust is myself, and if I didn't see it happening, then I don't believe it happened."*

However, this skepticism was not universal. In households with high shared visibility, where family members communicated frequently and saw each other daily, recordkeeping was viewed as straightforward rather than suspect. For these caregivers, brief, selective updates were sufficient because accountability was maintained through ongoing interaction rather than formal documentation.

*"I usually give my mom updates, and then they're usually pretty concise and to the point…it only includes the important…whatever stands out."*

Overall, caregivers favored tools that aggregated and visualized data across people and time, producing shared evidence rather than isolated reports. AI was welcomed insofar as it could synthesize and visualize this distributed accountability.

*"Um. If the summary report would be helpful, like, especially in scenarios where doctors and nurses are asking what time is she taking her medications? How often? What dose?"*

**Design implication:** AI verification systems should function as synthesizers of distributed records rather than substitutes for them, providing transparent, editable summaries that maintain collective visibility over caregiving evidence.

Together, these practices of detecting anomalies and assembling evidence set the stage for Theme 3, which examines how caregivers calibrate trust and control when working with AI-supported systems.

## 4.4 Theme 3: Conditional Trust in AI and Automation

Participants expressed **cautious optimism** about using generative AI for verification. Many viewed automation as a useful assistant for drafting questions, flagging anomalies, or summarizing patterns, but never as a final authority. Trust was conditional, negotiated, and adjustable.

*"AI can start the question first… then the caregiver just spends the time to understand what AI provided."*

This form of conditional trust aligns with the simulation's findings: contextual AI outputs improved usability, but caregivers demanded transparency and the ability to adjust system behavior.

#### 4.4.1 Transparency and Editability

Caregivers wanted to see how conclusions were reached and to modify system phrasing for clarity and empathy. Editable questions were seen as both usability and ethical features, essential for preserving dignity and ensuring cultural or linguistic fit.

*"Sometimes the wording isn't right… I'd want to change it so she understands."*

This expectation directly informed the CLTV principle of **human-led interpretability**, ensuring that caregivers can inspect and adjust AI-generated outputs.

#### 4.4.2 Adjustable Sensitivity and Thresholds

Caregivers requested control over alert thresholds, acknowledging that what counts as urgent varies across households and over time.

*"Having the ability to create a threshold… that baseline is going to shift over time. Something the family can decide and move as necessary."*

This echoes the simulation's flagging findings: accuracy was highest for safety-critical tasks but more variable for subjective ones. Participants emphasized that flexibility, rather than automation, is what builds sustainable trust.

#### 4.4.3 Reliability, Accountability, and Shared Oversight

Automation introduced new forms of vulnerability: when data failed to sync or anomalies were misclassified, caregivers felt increased anxiety. They stressed that the system's errors would still be their responsibility. As one caregiver explained, *"If you make the wrong guess as a caregiver, the healthcare system, or they make the wrong guess, they're not accountable. It's you that takes on the worries of the world that you made the wrong decision… But how do you make decisions like that? You need hard evidence."* This sense of residual accountability underscored the fragile trust between human and machine. Trust therefore depended on mutual accountability: the AI should make its reasoning transparent so caregivers can cross-check and correct it. Moreover, participants valued shared transparency across families, personal support workers (PSWs) and clinicians, transforming verification from a solitary burden into a collective practice.

However, not all participants shared these apprehensions. Some caregivers expressed confidence in the system's training and reasoning, emphasizing its dependability over the need for oversight. As one participant noted, *"Well, I guess you've got the AI rather trained in such a way it's gonna answer, do what it has learned. So it's okay with me."* Another described being comfortable with AI autonomy, remarking, *"Okay. Yeah. AI screening it. That's okay*." These views highlight how familiarity and trust in automation could mitigate concerns about accountability. Yet this confidence was largely anticipatory, reflecting participants' expectations before real-world use. Their opinions may evolve once they encounter the system's limitations and the practical realities of shared responsibility in caregiving contexts.

### 4.5 Cross-Theme Synthesis

Across themes, verification emerged as a **recursive feedback ecology** between caregivers and automation.

- **Anomalies** initiated attention.
- **Evidence construction** provided interpretive grounding.
- **Conditional trust** governed how caregivers calibrated system involvement.

These processes are cyclical rather than linear. When AI surfaces a new anomaly, caregivers evaluate it, contextualize it with physical and social evidence, and use their feedback to reshape how the system behaves in future iterations. This mutual adjustment defines the *caregiver-in-the-loop* dynamic that anchors the CLTV framework.

Verification, then, is not a technical audit but an *epistemic and emotional negotiation*. It involves balancing risk, trust, and care while interpreting uncertainty and maintaining relational integrity. Designing for this ecology requires tools that keep caregivers central, not peripheral, to algorithmic reasoning.

### 4.6 Transition to Framework

These findings directly informed the CLTV framework presented in the next section. CLTV formalizes the observed verification cycle into four design layers:

1. Anomaly Detection
2. Evidence Triangulation
3. AI-Assisted Summarization
4. Accountability Circulation

These results translate directly into CLTV's layered design for verification: Anomaly Detection (surface meaningful deviations), Evidence Triangulation (combine conversational + material traces), AI-Assisted Summarization (memory-light phrasing), and Accountability Circulation (shared, inspectable records), all under continuous Caregiver Oversight & Adjustment.

## 5 Caregiver-in-the-Loop Task Verification (CLTV) Framework

The Caregiver-in-the-Loop Task Verification (CLTV) framework conceptualizes task verification in dementia care as a continuous cycle of interpretation between caregivers and AI systems (Figure 2). At its core lies Caregiver Oversight and Adjustment, the active process through which caregivers interpret cues, fine-tune system behavior, and sustain ethical control. Surrounding this center are four interdependent processes: Anomaly Detection**,** Evidence Triangulation**,** AI-Assisted Summarization**,** and Accountability Circulation. Together, they form a sociotechnical loop in which automation supports human sensemaking but never replaces it.

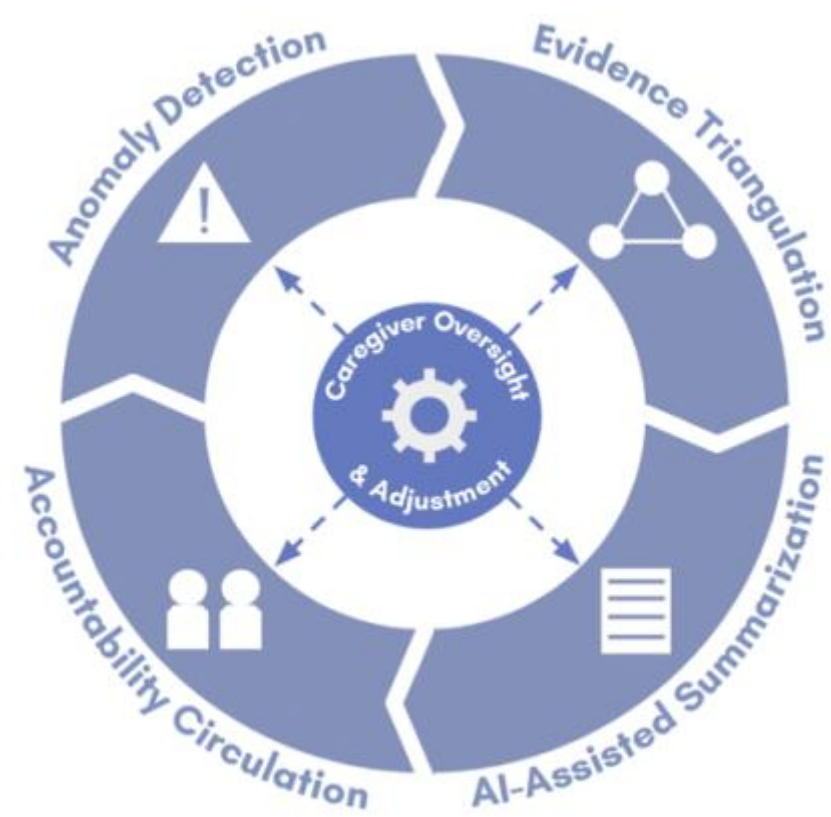

*Figure 2 Visual representation of the Caregiver-in-the-Loop Task Verification (CLTV) framework*

### 5.1 Anomaly Detection

Caregivers described verification as beginning not with confirmation but with noticing, recognizing when something seemed off. In both the interviews and the simulation probe, anomaly detection emerged as the entry point for reasoning. Systems that simply flagged every irregularity were seen as unhelpful; what caregivers needed were context-sensitive cues that prioritized meaningful deviations. They wanted to know when skipped meals or unsupervised outings required attention, not every trivial variation in routine.

AI can contribute by identifying such deviations, but the system must communicate them *as hypotheses*, not facts. In CLTV, anomaly detection represents the *system's attentional layer*: the AI surfaces patterns or absences it finds unusual and attaches confidence indicators and contextual notes explaining why they might matter. The caregiver then interprets whether this constitutes a true concern, maintaining ultimate judgment. This ensures that alerts prompt inquiry rather than anxiety, aligning automation with caregivers' interpretive rhythms.

### 5.2 Evidence Triangulation

Once an anomaly has been flagged, the process shifts from *noticing* to *substantiating*. Caregivers begin assembling evidence to confirm or dismiss what the system highlighted, rarely relying on a single signal alone. They combine environmental traces (a wet towel, an empty plate, a turned-off stove) with verbal exchanges and shared notes from other caregivers. Verification thus becomes an act of triangulation, where conversation, context, and material evidence are cross-checked to reconstruct what likely happened.

CLTV formalizes this practice by advocating transparency in how systems infer completion. Rather than simply reporting "task done," the system should explain why it inferred that outcome by showing which cues or patterns informed the conclusion. This enables caregivers to compare algorithmic reasoning with their own knowledge of the person's routines. Evidence triangulation therefore positions AI as a *sensemaking partner*: it gathers and organizes traces, while the caregiver performs the interpretive synthesis.

### 5.3 AI-Assisted Summarization

Across both phases of the study, caregivers emphasized the need for concise, interpretable summaries instead of raw data streams. The simulation probe illustrated how generative AI could assist by converting sensor and reminder information into short, memory-light statements, such as “Breakfast plate appears empty” rather than “Meal task marked completed.” Caregivers responded positively to this style of output, noting that it resembled how they already phrased supportive questions to the person living with dementia.

AI-Assisted Summarization in CLTV therefore serves as a bridge between detection and interpretation. It condenses multiple cues into a human-readable narrative that invites response and editing. Summaries must remain revisable: caregivers can adjust tone, simplify wording, or localize phrasing for cultural or emotional appropriateness. In this way, summarization becomes a conversational artifact that is jointly shaped by algorithmic processing and caregiver judgment.

### 5.4 Accountability Circulation

Verification rarely ends with a single individual. Caregiving is distributed across families, PSWs, and clinicians, all of whom contribute fragments of information about the person’s daily routines. Participants frequently described using whiteboards, group texts, or notebooks to coordinate tasks, but they also voiced frustration when these records became inconsistent or unverifiable.

Accountability Circulation captures how verification information moves through these networks. Within CLTV, AI-generated summaries and logs become *shared, editable artifacts* that circulate among caregivers and professionals. Each update records who made it and when, ensuring visibility without imposing blame. This transforms verification from a private burden into a collective accountability practice. By making records transparent and co-owned, systems can strengthen trust and coordination rather than replace them.

### 5.5 Caregiver Oversight and Adjustment

At the center of the framework is the caregiver’s ongoing interpretive work. Caregiver Oversight and Adjustment represents the continual calibration that keeps the verification loop contextual and humane. Caregivers review anomalies, interpret evidence, reword AI outputs, and modify thresholds to suit evolving needs. These adjustments feed back into the system, shaping how future anomalies are detected and summarized.

Rather than serving a purely supervisory role, oversight functions as an adaptive dialogue between caregiver and system. Through this process of co-learning, automation becomes increasingly aligned with the rhythms and values of the household. It ensures that accountability, tone, and interpretability remain anchored in lived caregiving practice.

### 5.6 Summary

The CLTV framework therefore models verification as an *interpretive partnership* between human and machine. Anomalies signal where attention is needed; evidence triangulation grounds that attention in context; AI-assisted summarization helps convert dispersed cues into understanding; and accountability circulation distributes that understanding across networks of care. At every stage, caregiver oversight closes the loop, adjusting both interpretation and automation.

By embedding human reasoning within algorithmic feedback cycles, CLTV reframes task verification as a process of collaborative sensemaking that continually negotiates evidence, risk, and trust. In doing so, it advances a model of human-AI collaboration that centers empathy, transparency, and shared responsibility in everyday dementia care. Although our examples use a large language model, CLTV is model-agnostic and can be implemented with rule-based logic, hybrid sensing pipelines, or other AI techniques, provided that editability, transparency, and caregiver control are preserved.

CLTV can be prototyped using multiple technical approaches. A large-language-model (LLM) pipeline can generate and summarize follow-up questions from reminder data, while a rule-based layer handles safety-critical checks (e.g., medication or stove use). Alternatively, a hybrid sensing interface could combine lightweight environmental cues (e.g., motion, temperature, object states) with caregiver annotations to triangulate evidence. Regardless of implementation, the framework requires an editable caregiver dashboard where thresholds, tone, and phrasing can be reviewed and adjusted, ensuring that interpretability and accountability remain grounded in caregiver control. Table 3 presents illustrative examples of how CLTV's layers could inform interface design and care practices. These are not prescriptive specifications but conceptual heuristics intended to demonstrate how interpretability, trust, and caregiver oversight might be operationalized in practice.

*Table 3 Illustrative design heuristics derived from the CLTV framework*

| CLTV Layer | Design Principle | Example Interface Feature | Potential Failure | Mitigation / Care Practice |
|---|---|---|---|---|
| **Anomaly Detection** | Context-sensitive surfacing | Adjustable "critical vs routine" filter with caregiver-defined priorities | Alert fatigue or desensitization | Sensitivity slider + narrative justification ("why flagged") |
| **Evidence Triangulation** | Transparent inference | "Why flagged" view showing which cues (e.g., towel sensor + verbal response) informed the result | False sense of certainty | Editable annotation + source links to raw evidence |
| **AI-Assisted Summarization** | Human-editable phrasing | Caregiver-editable prompt templates and tone presets | Tone mismatch or misinterpretation | Preview + tone adjustment controls |
| **Accountability Circulation** | Shared, traceable records | Multi-user log with edit history and contributor tags | Blame or surveillance anxiety | Collective visibility, non-punitive framing of edits |
| **Caregiver Oversight & Adjustment** | Continuous calibration | Dashboard for threshold, tone, and phrasing updates | Automation drift or misalignment | Periodic caregiver review prompts + learning history view |

# 6 Discussion

The CLTV framework extends conventional human-in-the-loop and mixed-initiative paradigms by shifting from supervisory override to interpretive calibration. Rather than positioning humans as final arbiters of algorithmic output, CLTV models how caregivers and AI systems co-adjust their reasoning through iterative feedback. This reframes control as an ongoing negotiation of meaning rather than a discrete act of validation, broadening existing theories of explainability, participatory AI, and co-adaptive automation within health-related HCI.

### 6.1 Verification as Relational Sensemaking

The CLTV framework reframes verification as a relational, situated process shaped by the emotional, ethical, and social dynamics of caregiving. Whereas many human–AI collaboration models, including human-in-the-loop and mixed-initiative systems, assume stable goals and well-defined decision criteria, dementia care unfolds through continual uncertainty and affective labor. Caregivers are not only checking whether a task was completed; they are simultaneously attending to wellbeing, dignity, safety, and the subtle cues that signal changes in a person's cognitive or emotional state. Their reasoning therefore operates on intertwined practical and relational levels, integrating AI outputs into an ongoing process of interpretation, reassurance, and moral accountability.

Our findings extend longstanding HCI critiques of conversational agents as unpredictable, ambiguous, and socially fragile systems (Luger and Sellen, 2016; Porcheron et al., 2018; Pradhan et al., 2018). Caregivers in our study encountered many of these same challenges, including variable tone, inconsistent interpretations, and moments of friction, but within a context where such breakdowns carry heightened emotional weight. Rather than relying on AI responses at face value, caregivers engaged in rich interpretive work, triangulating AI outputs with contextual cues, lived experience, and embodied knowledge of the person they support. This aligns with HCI research showing that users navigate AI ambiguity through ongoing interpretive negotiation (Ehsan et al., 2021), but it extends this literature by demonstrating how such negotiation becomes inseparable from the relational responsibilities of care.

HCI scholarship on care ethics similarly emphasizes that caregiving decisions are socially and emotionally situated, and that technologies must support, rather than flatten, these interpretive practices (Berridge et al., 2021; Lazar et al., 2017). CLTV reflects this orientation by treating verification not as a binary judgment but as an ongoing, relational process of understanding and negotiation. In this framing, explanation in AI shifts from a primarily epistemic mechanism ("why did the system decide this?") to a dialogic and relational one ("does this reasoning make sense here, with this person, in this moment?"). This broadens the scope of explainable and trustworthy AI by emphasizing meaning-making, relational alignment, and the situated interpretive strategies that caregivers develop as part of everyday care.

### 6.2 Extending Human-in-the-Loop to Caregiver-in-the-Loop

The notion of keeping caregivers "in the loop" expands conventional human-in-the-loop paradigms. In many HCI and AI systems, the human functions as an occasional validator or corrective mechanism, stepping in when automation misfires. By contrast, the caregivers in this study occupied a far more constitutive role: their interpretations continuously shaped what the AI attended to, how the system phrased its questions, and which events warranted escalation. The loop was recursive rather than supervisory, as caregiver reasoning was not simply reacting to the system's outputs but actively co-constructing them. Over time, this feedback informally tuned anomaly thresholds, shaped linguistic tone, and influenced how verification unfolded. This reframing aligns with HCI research on mixed-initiative interaction, explainable AI, and trust calibration, which similarly emphasize systems that support users as active sensemakers rather than passive recipients of output (Amershi et al., 2019; Eiband et al., 2018; Eric Horvitz, 1999). Yet CLTV adapts these principles to a domestic care environment where emotional labor, relational expertise, and lived knowledge are central. Unlike traditional human-in-the-loop settings, which are often oriented around professional operators and bounded decision criteria, caregiver-in-the-loop collaboration is situated, affective, and improvisational. It requires technologies to accommodate uncertainty, relational nuance, and the moral texture of everyday care decisions.

Our findings also resonate with emerging HCI scholarship that conceptualizes LLMs as generative design materials rather than autonomous agents (Mao et al., 2023). Caregivers' iterative guidance of the AI system parallels this vision of co-adaptation: automation learns not only from data, but also from caregivers' contextual judgments and corrections; caregivers, in turn, learn how to interpret and shape algorithmic cues. This mutual adaptation recasts control not as a binary allocation between human and machine, but as an ongoing process of calibration across difference, between formal model logic and the embodied expertise of care.

The CLTV framework thus contributes a care-centered model of human–AI collaboration. Future design work should operationalize this reciprocity by creating interface mechanisms that allow caregivers to teach and adjust the system (for example, by editing phrasing, modifying thresholds, or viewing and revising the system's reasoning traces). Such mechanisms can support the relational and interpretive work that is foundational to dementia care, while grounding AI behavior in the values and expertise of those who provide care.

### 6.3 Ethics and Emotional Labor in AI-Mediated Care

Designers should therefore evaluate emotional as well as functional outcomes when assessing system success. Even when automation contributed to a decision, caregivers in our study remained accountable for its consequences, reflecting what feminist and care-ethics scholars describe as the moral stickiness of delegated responsibility. The CLTV framework addresses this complexity by foregrounding accountability circulation: verification becomes a visible, shareable practice distributed across the care network, diffusing responsibility while preserving empathy and relational sensitivity.

HCI work on domestic technologies has shown that seemingly small interface decisions can subtly shift responsibility, influence feelings of being supported or blamed, and shape how family members interact with one another (Odom et al., 2012). Our findings extend this line of research by illustrating how AI-generated cues become part of a broader ecology of accountability within households, sometimes redistributing interpretive labor, sometimes amplifying emotional load, and sometimes offering relief. These dynamics reinforce longstanding HCI arguments that domestic technologies do not merely stabilize existing relationships but actively participate in renegotiating roles, responsibilities, and perceptions of care (Elsden et al., 2016; Neustaedter and Sengers, 2012).

This reframing has important implications for responsible AI design. Ethical alignment cannot be achieved solely through privacy protections or fairness metrics; it must also encompass affective alignment, meaning the system's ability to respect caregivers' emotional rhythms, interpersonal values, and relational modes of reasoning. Designing for CLTV therefore means designing for trust maintenance, not only trust calibration: the system should help caregivers feel informed, supported, and in control, even when uncertainty is inherent and outcomes cannot be fully predicted.

### 6.4 Design Implications for Everyday AI Systems

Translating CLTV into practice requires interaction patterns that make reasoning both visible and adaptable. Interfaces should present anomalies as narrative cues, not numeric scores. Summaries should show why evidence was combined, with options to rephrase or annotate outputs. Thresholds must be adjustable and explainable, allowing families to calibrate sensitivity over time.

In practice, this implies guardrails such as confidence displays, fallback phrasing for ambiguous outputs, and human-editable templates to mitigate variability in generative models. Beyond dementia care, these insights also apply to domestic and health-support AI contexts where automation intersects with moral responsibility, including areas such as elder-care robotics, medication adherence, and home monitoring. CLTV demonstrates how explainability and adaptability can be embedded in *everyday AI*, translating abstract principles of transparency and control into tangible interaction design choices.

### 6.5 Methodological Reflections

Methodologically, the study demonstrates the value of combining speculative simulation with qualitative inquiry. The simulation probe acted as a boundary object, provoking concrete reflection and enabling participants to articulate values that might otherwise remain tacit. Using generative AI to produce examples made abstract issues of verification tangible, bridging technical feasibility and experiential insight. This approach aligns with emerging critical-speculative HCI methods that treat AI prototypes not as solutions but as thinking tools for examining sociotechnical futures.

At the same time, the study's dual role in simulation development and interview analysis calls for reflexivity. The first author's technical background facilitated translation between caregivers' reasoning and AI implementation but risked bias toward computational framing. Continuous consultation with lived-experience advisors helped ensure interpretive resonance, reminding that co-design with caregivers must prioritize relational knowledge as much as algorithmic performance.

### 6.6 Limitations and Future Directions

While CLTV provides a conceptual foundation, it has not yet been validated in deployed systems. Future work should operationalize its principles through iterative prototyping and field evaluation, measuring not only task accuracy but also caregivers' cognitive load, perceived trust, and emotional wellbeing. Broader sampling that includes diverse cultural contexts, stages of dementia, and care network configurations would strengthen generalizability. Technical exploration is also needed to translate caregiver feedback loops into machine-learning architectures capable of dynamic, user-driven calibration. Generative models can exhibit variability and drift across updates; deployed systems will require versioning, confidence indicators, and safe fallbacks (e.g., default templates) to maintain consistency and avoid brittle behavior. These directions point toward concrete system-design experiments that iteratively prototype caregiver-in-the-loop interfaces and measure their interpretive usefulness in real homes.

More broadly, CLTV opens space for cross-disciplinary collaboration. Integrating this framework with explainable AI research, participatory design, and cognitive assistive technology can move the field toward co-adaptive care systems that learn from relational expertise rather than replacing it. Finally, CLTV is a framework, not a single algorithm; future implementations could combine LLMs with lightweight sensing or rule-based checks to balance interpretability, cost, and robustness.

### 6.7 Summary

Through the CLTV framework, this study advances a model of human-AI collaboration grounded in the lived epistemology of care. Verification emerges as a recursive process of detection, interpretation, and negotiation, sustained by caregiver oversight and distributed accountability. By making reasoning transparent, editable, and socially visible, CLTV repositions automation as a participant in relational sensemaking. It contributes to the broader discourse on responsible AI by demonstrating that interpretability, trust, and ethics must be designed not only into algorithms but into the *relationships* they inhabit. Beyond dementia care, the framework's emphasis on interpretive calibration and shared accountability can inform AI support for elder-care, medication adherence, rehabilitation tracking, and home-monitoring contexts where relational reasoning is central.

## 7 Conclusion

This study examined how caregivers conceptualize task verification within dementia care and how generative AI might support, rather than replace, their interpretive reasoning. Through qualitative interviews and a speculative simulation probe, we found that verification is less a matter of checking compliance than of constructing trustworthy evidence under uncertainty. Caregivers engage in continuous sensemaking by detecting anomalies, interpreting contextual cues, and calibrating trust across human and technological actors.

The proposed Caregiver-in-the-Loop Task Verification (CLTV) framework formalizes this process as a sociotechnical cycle of *anomaly detection, evidence triangulation, AI-assisted summarization,* and *accountability circulation*, centered on caregiver oversight and adjustment. CLTV advances current understandings of human-AI collaboration by demonstrating how interpretability and control can be distributed across people and systems in relational settings. Rather than framing AI as an autonomous verifier, it positions automation as a conversational and evidential partner that helps caregivers manage uncertainty, reduce cognitive load, and maintain moral accountability. While developed for dementia care, CLTV's principles extend to other domestic health contexts where verification is interpretive and relational.

By integrating technical speculation with lived experience, this work extends design theory for responsible AI in health and domestic contexts. It invites future research to translate CLTV principles into deployable systems and to explore how caregiver-in-the-loop verification might generalize to other domains where human judgment and ethical responsibility intertwine. Ultimately, the study argues that advancing AI in caregiving requires not only better models of intelligence but also better models of relationship, with systems designed to think with caregivers rather than for them.

**Funding sources**

This work was supported by the University of Toronto, AGE-WELL NCE, and the Canadian Institutes of Health Research.

**Declaration of generative AI use**

During the preparation of this work the authors used GPT-4 to generate example follow-up questions and simulated responses for a formative simulation probe and to assist with wording suggestions during drafting. After using this tool, the authors reviewed and edited all content and take full responsibility for the final manuscript. No figures were created or altered using generative AI.

## Appendix A: Evaluation Criteria Checklist

| Criteria | Questions for Evaluation |
|---|---|
| Clarity | Is the question easy to understand without ambiguity or overly complex phrasing? |
| Conciseness | Does the question communicate its purpose in as few words as necessary while maintaining full meaning? |
| Contextual Relevance | Does the question refer specifically to the task or details mentioned in the reminder? |
| Avoidance of Assumptions | Does the question avoid introducing or assuming details that were not part of the original reminder? |
| Memory Independence | Does the question include enough contextual clues (e.g., time, place, or actions) to minimize reliance on the user's memory? |
| Support for Recall | Does the question avoid requiring the user to recall details without support or prompts? |
| Reminder Specificity | Could a more detailed or specific reminder make the question clearer or more actionable? |
| Task Completion Focus | Is the question focused on determining whether the task from the reminder was completed? |
| Avoidance of Irrelevance | Does the question avoid straying into unrelated or irrelevant topics? |
| Supportive Tone | Does the question use a supportive and respectful tone? |
| Encouraging Engagement | Does the tone encourage engagement without being judgmental or condescending? |

## Appendix B: Decision tree for response classification

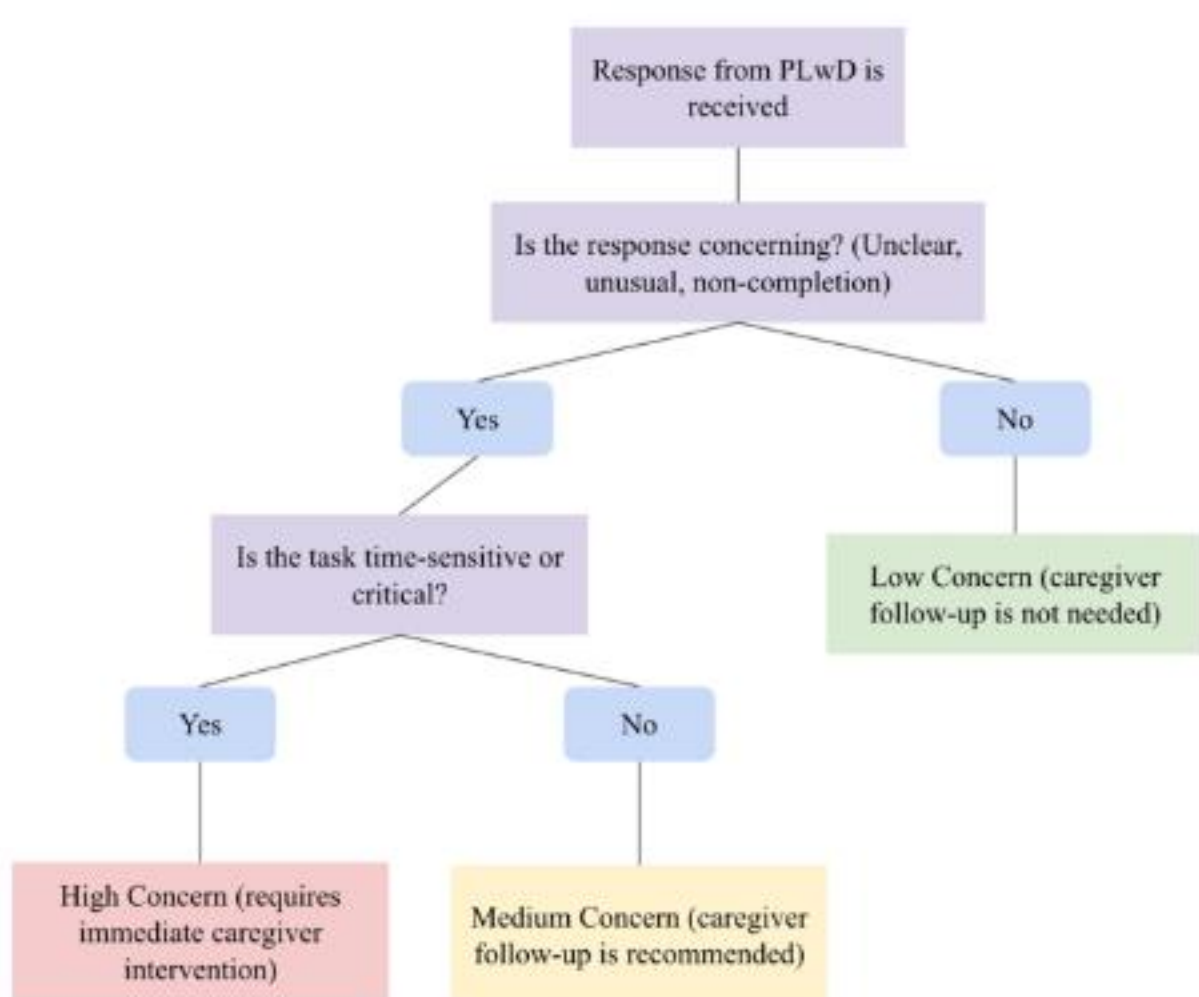